\documentclass{webofc}
\usepackage[varg]{txfonts}   
\begin{document}
%
\title{Studying short-range nuclear correlations using relativistic heavy-ion collisions}
%
%

\author{\firstname{Matthew} \lastname{Luzum}\inst{1}\fnsep\thanks{\email{mluzum@usp.br}} \and
        \firstname{João Paulo} \lastname{Picchetti}\inst{1}\fnsep
        \and
        \firstname{Mauricio} \lastname{Hippert}\inst{2}\fnsep
             \and
        \firstname{Jean-Yves} \lastname{Ollitrault}\inst{3}\fnsep
}

\institute{
Instituto de Física, Universidade de São Paulo,
Rua do Matão, 1371, 05508-090, São Paulo, Brazil
\and
           Illinois Center for Advanced Studies of the Universe  \&
           Department of Physics, University of Illinois Urbana-Champaign, 1110 W. Green St.,
           Urbana IL 61801-3080, USA.
\and
           Université Paris Saclay, CNRS, CEA, Institut de physique théorique, 91191 Gif-sur-Yvette, France
          }

\abstract{%
  Recently, a method was developed for implementing arbitrary short-range nucleon-nucleon correlations in Monte Carlo sampled nuclei (as well as deformations of the 1-body nuclear density).   We use this method to implement realistic 2-body correlations in a sample of nuclei for use  in simulations of relativistic heavy-ion collisions and we quantify the statistical benefits.   These results demonstrate that the method can be used to easily implement an arbitrary correlation function, and systematically study the effects of correlations using significantly less resources than is necessary with traditional methods.
}
\maketitle
\section{Introduction: Nuclear structure via shifting nucleons}
\label{intro}
There is increasing interest in studying the effects of low-energy nuclear structure on high-energy nuclear collisions.  These properties include shape deformations, which encode information about long-range correlations in the nucleus.   However, short-range interactions between nuclei (in addition to Pauli repulsion) ensure that there are also important short-range nucleon-nucleon correlations.   These correlations have been studied in the context of heavy-ion collisions \cite{Alvioli:2009ab}, but the effects are still not well understood.  Typically in modern simulations, these effects are crudely mimicked by implementing in some way an exclusion radius --- that is, two nucleons are not allowed to be closer than some minimum distance. (Meanwhile the correlation at larger distances is either maintained close to zero, or left uncontrolled).  Depending on the implementation, even the 1-body density can be inadvertently modified when ensuring the exclusion distance is respected.

Recently, methods were introduced for implementing and systematically studying such nuclear structure effects in simulations \cite{Luzum:2023gwy}.  
By making small changes in the positions of nucleons in a collection of nuclei, one can obtain nuclei with any desired properties, but with a statistical uncertainty that is strongly correlated between each collection.  As such, there is strong cancellation in statistical uncertainties of \textit{relative} quantities, and one can very accurately map the effect of changing nuclear structure with significantly reduced resources.

In addition to the computational benefits, these methods also have the advantage that the 1-body and 2-body densities can be directly controlled, and respecting arbitrary desired functions (so long as the 2-body correlations are sufficiently ``short-range'' in comparison to the system size).    An implementation of these methods is publicly available \cite{code}

\section{Implementation of realistic 2-body correlation}

In reference \cite{Alvioli:2009ab}, methods were developed to generate nuclear configurations that include effects of realistic 2-body and 3-body interactions, and these methods were used in reference \cite{Hammelmann:2019vwd} to generate 10000 configurations of $^{96}$Ru nuclei (relevant, for example,  for the study of Ru and Zr isobar collisions that were performed at RHIC).


\begin{figure}
\centering
\sidecaption
\includegraphics[width=6cm,clip]{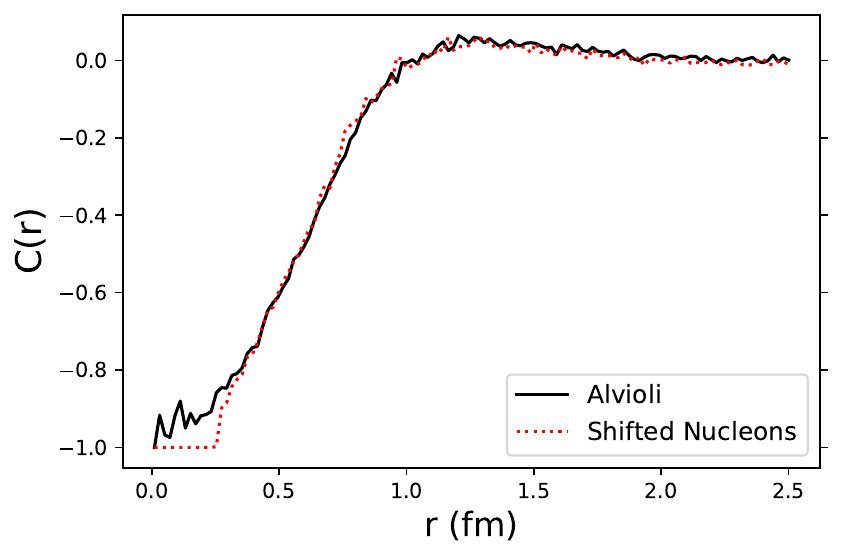}
\caption{Correlation function extracted numerically from Monte-Carlo-sampled $^{96}$Ru nuclei as prepared in \cite{Hammelmann:2019vwd} (Alvioli) and using the methods of this work \cite{Luzum:2023gwy} (Shifted Nucleons)}
\label{fig:corr}       
\end{figure}

To illustrate the benefits of our methods  \cite{Luzum:2023gwy}, we numerically extract the (angle-averaged) 2-body correlation function of these sampled configurations,  
\begin{align*}
C(r) &= 
\left\langle \frac {\rho_2(\vec x_1, \vec x_2)}{\rho(\vec x_1) \rho(\vec x_2)}\right\rangle - 1
\simeq
\frac{\int d\bar x d\Omega\, \rho_2(\vec x_1, \vec x_2)}{\int d\bar x d\Omega\, \rho(\vec x_1) \rho(\vec x_2)} - 1
\end{align*}
with $\rho_2$ the 2-body probability density, $\rho$ the 1-body density,  $\bar x = (\vec{x_1}+\vec{x_2})/2$, and $r = |\vec{x_1} - \vec{x_2}|$.
The resulting correlation function as shown in Fig~\ref{fig:corr} was used as input to the shifting-nucleon algorithm, and new configurations generated that respect this correlation function along with the same Woods-Saxon 1-body distribution ($R$ = 5.085 fm, $a$ = 0.46 fm,  $\beta_2$ = 0.158), shown also in Fig~\ref{fig:corr}.\footnote{We note that the numerical uncertainty in the extracted correlation at small $r$ (only 10000 nuclei were generated for \cite{Hammelmann:2019vwd}, and few pairs exist at small $r$)  affects the numerical determination of nucleon shifts and causes the small deviation  for $r<0.3$ fm.  Manually smoothing the data could improve accuracy, but has not been done here.}
Details of the method and implementation can be found in \cite{Luzum:2023gwy}, and the resulting code is available for use \cite{code}.

We note that the resulting nuclei are not identical to those from \cite{Hammelmann:2019vwd}, in particular because of the lack of 3-body correlations.   So while our method is superior to those typically used in simulations, it is inferior to this more sophisticated method in that sense.  However, it still retains significant advantages with regard to public availability, ease-of-use, and especially the ability for systematic variation and efficient study of the effects on collision observables.

We illustrate these benefits via a simple example Bayesian analysis in the following.

\section{Effect on heavy-ion collisions}

To illustrate the methods, we set up central ($b$=0) collisions between nuclei with varying short-range correlation functions, using the popular TrENTO model \cite{Moreland:2014oya} ($p$=0) to describe the distribution of energy in the transverse plane at early times.  We compute eccentricities $\varepsilon_n\{2\}$, which are closely related to final-state flow observables $v_n\{2\}$.

\subsection{Efficiency gain}
\begin{figure}
\centering
\includegraphics[width=0.9\linewidth,clip]{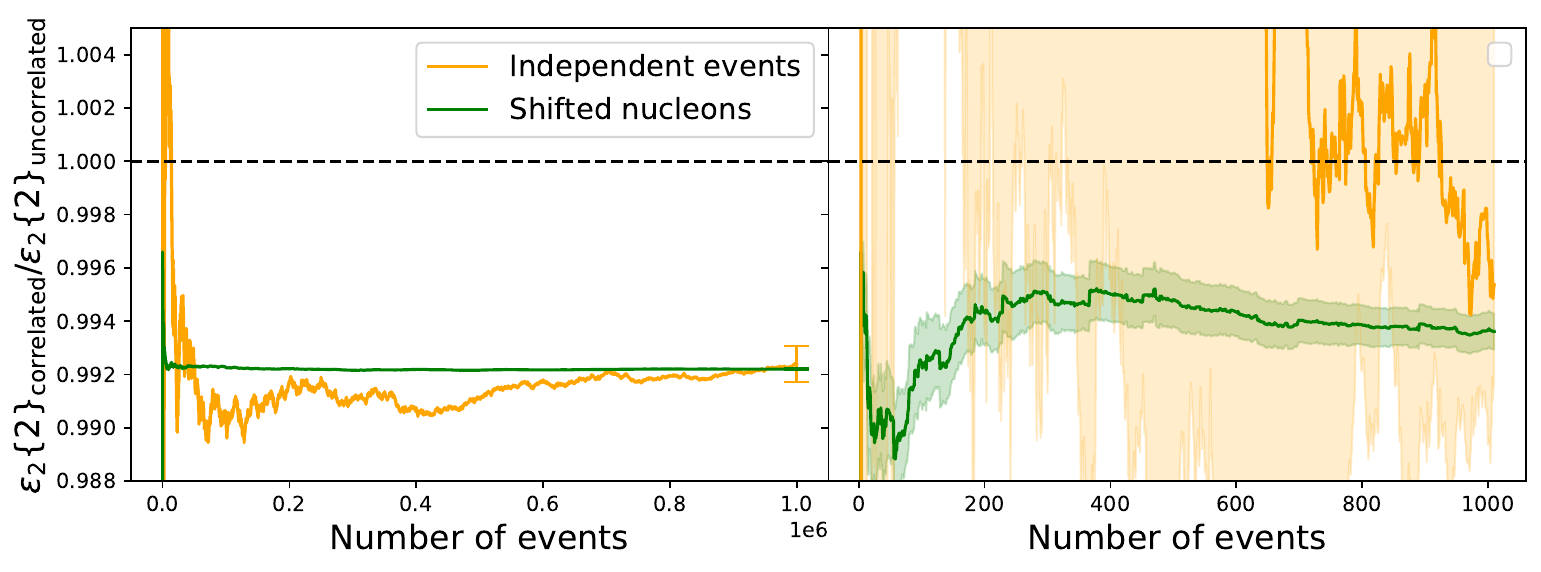}
\caption{Relative eccentricity of collision system with realistic nucleon-nucleon correlations (implemented as described above) compared to uncorrelated nucleons $\varepsilon_n\{2\}_{\rm correlated}/\varepsilon_n\{2\}_{\rm uncorrelated}$.  The $x$ axis shows an increasing number of collision events included in the calculation.  On the right plot the $x$ axis is zoomed in to show that with $\sim$1010 events such that the correlated and uncorrelated set are related via shifting nucleon positions, the uncertainty is approximately the same as 1000000 events when nuclei are prepared independently, saving almost 3 orders of magnitude computing time.
}
\label{fig:e22}       
\end{figure}

Here we compare the statistical uncertainty on the relative change in eccentricity when moving from uncorrelated nucleons to the realistic correlation function shown in Fig.~\ref{fig:corr}, $\varepsilon_2\{2\}_{\rm correlated}/\varepsilon_2\{2\}_{\rm uncorrelated}$.  From Fig.~\ref{fig:e22} we can see that the change in eccentricity is slightly less than 1\%.  
With more events, the statistical uncertainty (shown as a shaded band on the right plot and an error bar on the last point of the left plot) decreases.  
In order to resolve this small change using traditional methods (with every nucleus prepared independently), it requires quite a large number of events. However, 
if nuclei are prepared such that the correlated set and uncorrelated set are related via shifting nucleons, the statistical uncertainty largely cancels in the ratio, and the statistical demands are dramatically reduced.  In this particular case, the statistical uncertainty for 1 million events using the standard method is approximately the same as with our improved method with only 1010 events --- a speedup of almost 3 orders of magnitude.
While the quantitative benefit varies with the context \cite{Luzum:2023gwy} (generally, when the change to the nucleus is smaller, the relative efficiency gain is greater), it is clear that the potential benefits can be dramatic.


\subsection{Application to Bayesian analysis}
As an illustrated application to a modern Bayesian analysis, we consider a simple parameterized step-function correlation function, with variable strength $C_{\rm str}$ and length $C_{\rm len}$,
 $C(r) = C_{\rm str}\ \ \Theta(C_{\rm len} - r)$

\begin{figure}
\centering
\includegraphics[width=0.49\linewidth,clip]{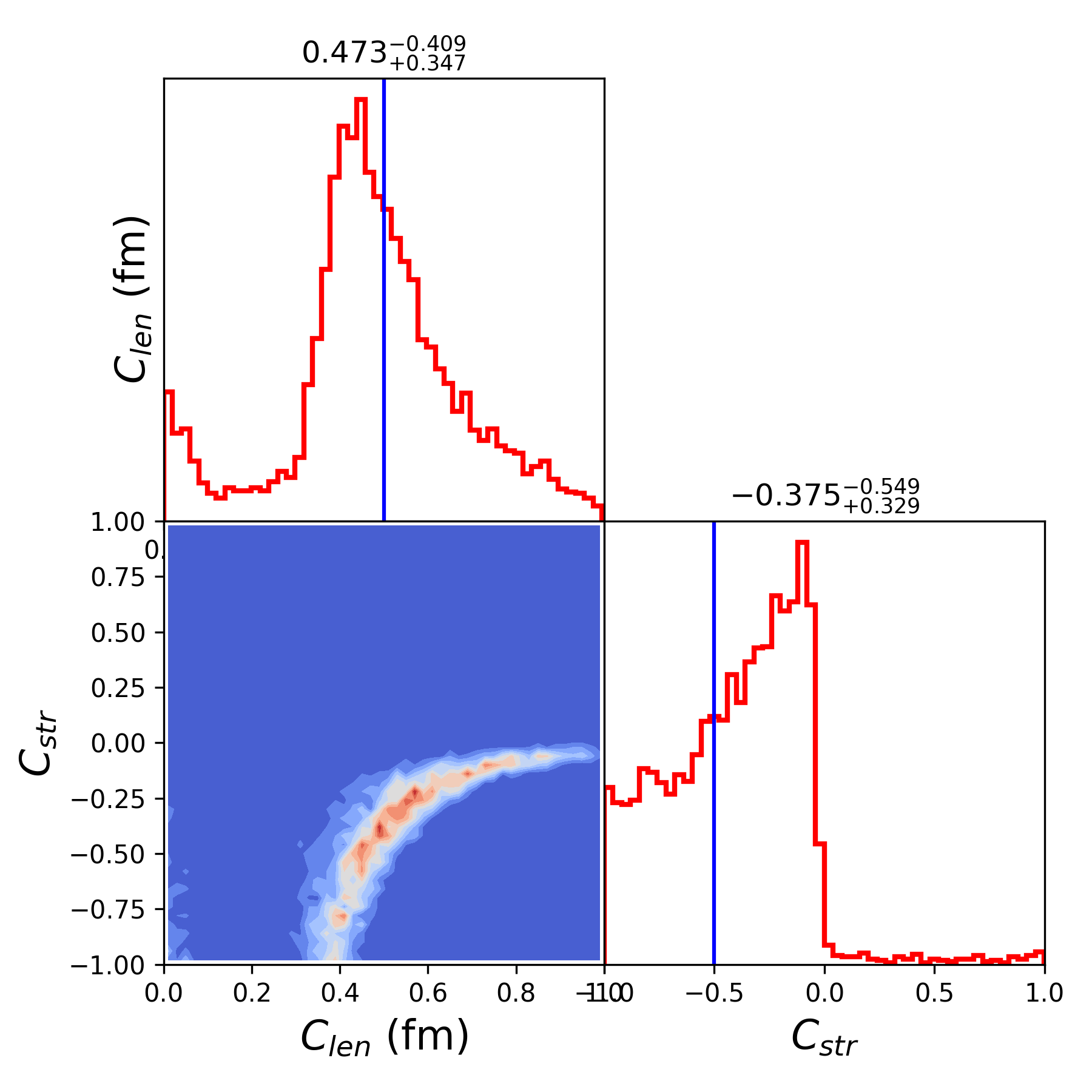}
\includegraphics[width=0.49\linewidth,clip]{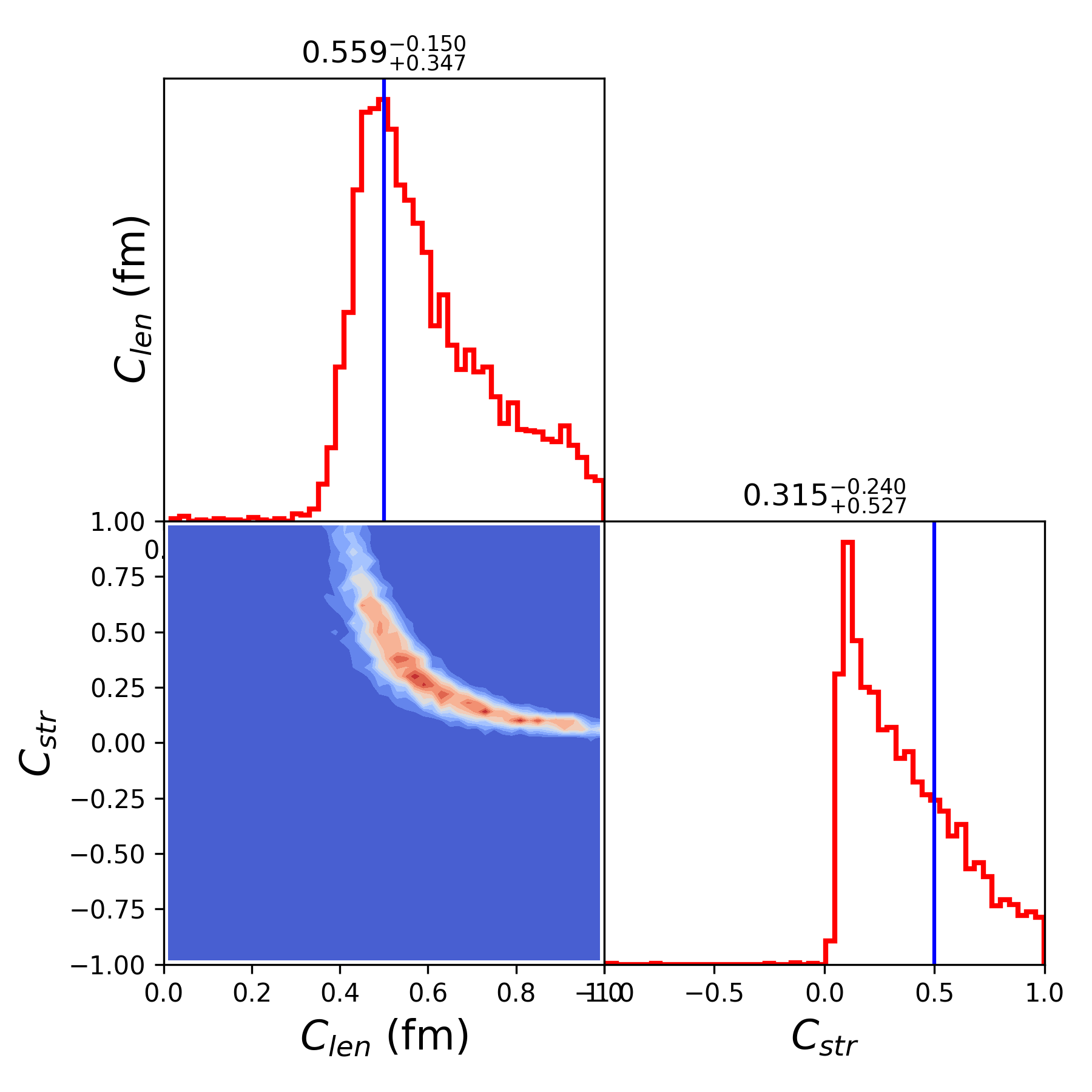}
\caption{Bayesian posterior probability distribution for a parameterized step-function correlation 
for two closure tests.  On the left, the model output for $C_{\rm len} = 0.5$ fm and $C_{\rm str} = -0.5$ was used as pseudodata.  On the right is shown a closure test for a hypothetical attractive potential $C_{\rm len} = 0.5$ fm and $C_{\rm str} = 0.5$. 
}
\label{fig:posterior}       
\end{figure}

We perform a closure test on this model --- that is, we generate model output $\varepsilon_n\{2\}$ for $n = 2,3,4,5$, and use them as pseudodata for Bayesian inference.  The resulting posterior distribution (shown in Fig.~\ref{fig:posterior}) should be consistent with the input parameter values (shown as vertical lines).  The results show not only the potential sensitivity of these particular observables on the parameters, but also redundancies and correlations between parameters.  
Here, one can see that a change in correlation length can be largely compensated by a change in strength, so that additional observables would be necessary to discern details of the shape of the correlation function.

\section{Summary}
With newly-developed methods for implementing short-range nucleon-nucleon correlations (in addition to shape deformations) in Monte Carlo sampled nuclei, we illustrate how an arbitrary correlation function can be implemented (including a realistic numerical correlation extracted from previously-generated nuclear configurations).   We also show that these methods can dramatically reduce computational demands when investigating the effects of nuclear properties on high-energy collisions.  

These developments open a vast set of possibilities for systematic studies of nuclear structure in ultrarelativistic heavy-ion collisions.  As an illustration of this, we show a simple Bayesian parameter estimation of the short-range nucleon nucleon correlation using early-time eccentricities.

\section*{Acknowledgements}

This work was supported in part by the National Science Foundation (NSF) within
the framework of the MUSES collaboration, under grant
number No. OAC-2103680; 
by the São Paulo Research Foundation (FAPESP) under grants  2018/24720-6, and 2017/05685-2; and by the Brazilian National Council for Scientific and Technological Development (CNPq). 

\end{document}